# User Satisfaction: UX Design Strategies for Seamless Virtual Experience

*Harish Vijayakumar, Design Solutions Engineer*
*Tamil Nadu, India. Email: thisisharish01@gmail.com*

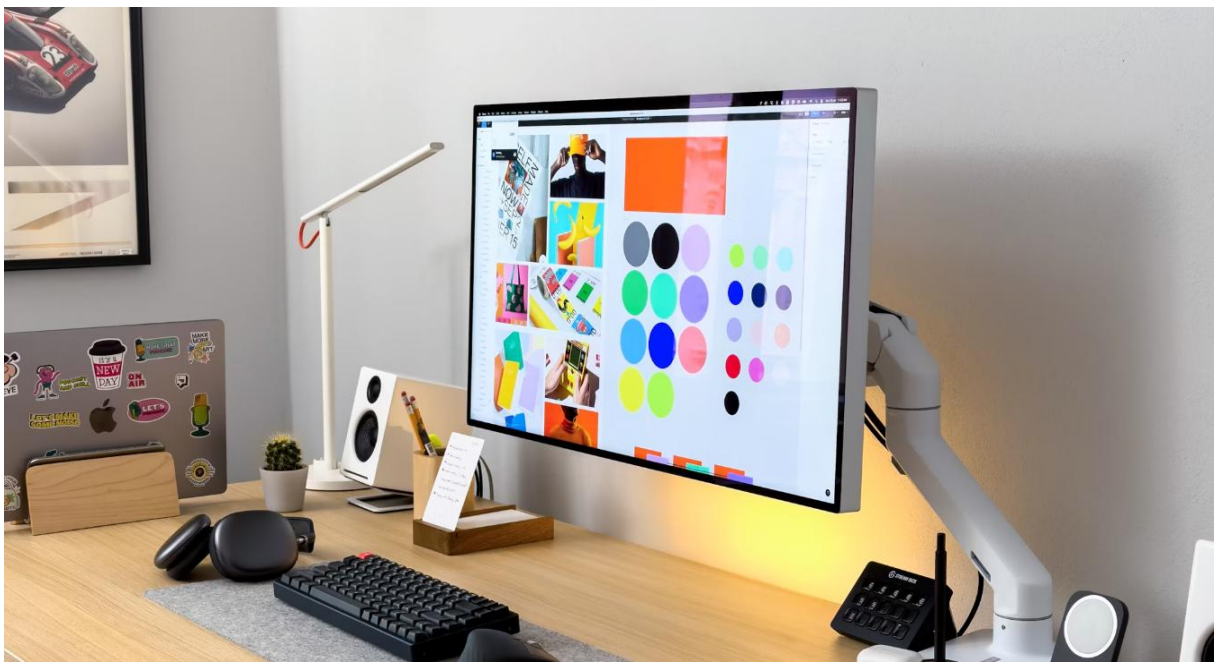

## Abstract

User Experience (UX) in virtual worlds is a fast-developing discipline that requires creative design concepts to overcome the divide between physical and virtual interaction. This research investigates primary principles and techniques to improve UX in virtual experiences based on usability, accessibility, user engagement, and technology advancements. It gives detailed insight into trends, issues, and prospects for UX design of virtual applications that guarantee an efficient, easy-to-use, and immersive experience.

## 1. Introduction

Virtual experiences transformed digital interactions across various sectors ranging from gaming and entertainment to learning and e-commerce. A strategically designed UX plan is critical for making users efficient in navigating and interacting with virtual environments. Several UX principles enabling smooth virtual experiences are discussed within this paper and highlighted are the usability, accessibility, and user-centred design considerations. The study examines how recent technological developments like augmented reality (AR), virtual reality (VR), and artificial intelligence (AI) affect UX design in virtual worlds.

## 2. Understanding User Needs in Virtual Experiences

Design for virtual experiences involves in-depth knowledge of user behaviour and expectations. Users want intuitive navigation, realistic interaction, and low cognitive load while in virtual environments.

To attain this, UX designers need to carry out user research in the form of surveys, usability testing, and behaviour analytics. User personas and journey mapping are important in understanding pain points and areas for improvement in virtual environments.

Creating Emotional Connections in Virtual Spaces: Beyond functionality, the way users feel in a virtual environment significantly impacts their overall experience. When a space feels immersive and natural, users are more likely to stay engaged. Simple elements like smooth animations, responsive interactions, and adaptive lighting can make a digital world feel more believable. The goal is to reduce any sense of detachment and create an environment that feels intuitive and alive.

Giving Users a Sense of Control: A great virtual experience doesn't just guide users also, it empowers them. When users feel they can influence their surroundings, they develop a stronger connection to the space. Allowing meaningful choices, personal customization, and interactive elements fosters a sense of agency. A well-designed virtual experience ensures users aren't just passive observers but active participants, making their journey feel personal and engaging.

## 3. Usability Principles for Virtual Environments

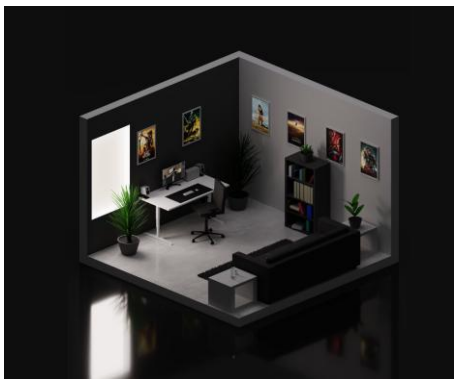
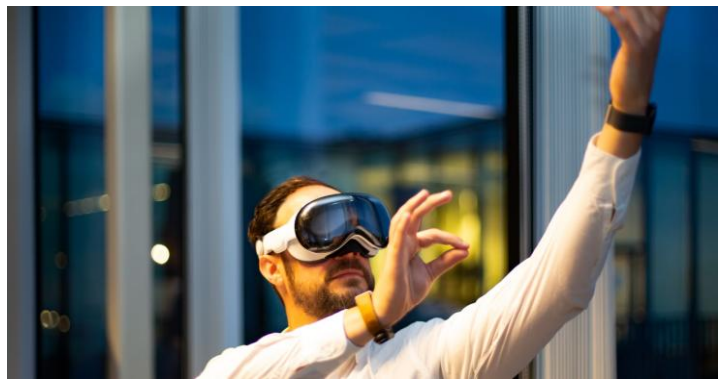

Usability continues to be a fundamental tenet of UX design. For virtual experiences, designers need to facilitate ease of use through effective navigation cues, responsive controls, and adaptive interfaces. Consistency across interfaces, minimalist design, and intuitive patterns of interaction diminish cognitive load and improve user satisfaction. Incorporating real-time feedback mechanisms like haptic feedback and visual cues enhances usability and creates engagement.

The principles to be followed when a user experience is being made or updated are mentioned to create visual experience seamlessly and give a better experience.

- Keep Things Consistent – Users should feel at home no matter where they are in the virtual space. Consistent controls, navigation, and interactions help reduce confusion and make the experience feel natural.
- Reduce Lag, Keep It Smooth – Nothing breaks immersion faster than delays. A responsive system with minimal latency ensures users stay engaged and don't get frustrated with sluggish performance.
- Make Navigation Intuitive – Users shouldn't have to struggle to move around. Virtual environments should mimic real-world movement where possible or use clear and simple control schemes that feel instinctive.
- Think About Everyone – Accessibility is key. Features like voice controls, customizable interfaces, and alternative navigation options make virtual spaces more inclusive for people with different abilities.
- Prevent Mistakes and Make Fixes Easy – No one likes being stuck after making a mistake. Virtual experiences should have clear error messages, undo options, and intuitive ways to get back on track.
- Use Realistic Feedback – Whether it's visual cues, sounds, or even haptic feedback, users need confirmation that their actions are being recognized. A slight vibration when pressing a button or a soft glow when selecting an object makes interactions feel real.
- Make It Work on Any Device – A good virtual experience shouldn't be limited to high-end devices. Designing for multiple platforms, from VR headsets to desktop screens, ensures more people can access and enjoy it.
- Let Users Personalize Their Experience – Not everyone interacts with technology the same way. Providing options to adjust brightness, control sensitivity, or even UI layouts can make the experience more comfortable.
- Don t Overload the Brain – Too much information at once can be overwhelming. Breaking content into digestible chunks and highlighting the most important elements helps users stay focused and engaged.

- Encourage Collaboration – Whether it's for work, gaming, or socializing, virtual spaces are often about connecting with others. Smooth communication tools, shared interaction mechanics, and synchronized environments help make collaboration effortless.

## 4. Addressing Challenges in Virtual UX Design

In spite of strides in UX design for virtual environments, problems like motion sickness, technical, and hardware constraints continue to exist. Designers have to counteract these challenges by providing optimized performance, ergonomic design, and responsive feedback systems. Testing and iterative development cycles ensure that user issues are catered to and virtual experiences are improved for more usability and comfort.

## 5. Accessibility in Virtual UX Design

Provision of accessibility in virtual worlds is paramount to enabling an inclusive experience for everyone. Designers need to address aspects of colour contrast, text readability, voice instructions, and alternative navigation options for differently-abled people. Compliance with accessibility standards such as the Web Content Accessibility Guidelines ([WCAG](#)) guarantees that virtual applications support different user needs. Inclusive design principles enhance usability and extend the reach of virtual experiences.

Let's breakdown the User Satisfaction with a formula based on which the exact need could be sorted out with ease.

$$User\ Satisfaction, US = \alpha(2EN + IQ) + \beta(2LTE + VC)$$

Where,

- US -> User Satisfaction
- EN -> Engagement
- IQ -> Interface Quality
- LTE -> Low Technical Errors
- VC -> Visual Comfort
- α   -> Weight for Engagement & Interface Quality
- β   -> Weight for Low Technical Errors & Visual Comfort

We could derive it from its origin:

Step 1: First Component Calculation:

$$U1 = 2EN + IQ$$

Step 2: Second Component Calculation:

$$U2 = 2LTE + VC$$

Step 3: Weighted Contribution

$$US = \alpha U1 + \beta U2$$

Step 4: Scaling to a 0-100 Range:

$$US(scaled) = \left[\frac{US}{MaxValue}\right] \times 100$$

The proposed User Satisfaction (US) formula provides a structured approach to quantifying UX effectiveness in virtual environments by integrating navigation ease, interaction quality, load time efficiency, and visual comfort.

By normalizing the results, this formula ensures a scalable and comparative metric that can be adapted for diverse digital experiences, facilitating data-driven UX improvements.

To have a better understanding, the following real-world experiment: User Satisfaction for [Apple Vision Pro](#) conducted would give an enhanced understanding and gives the desired value from the User Satisfaction rate which we are expected to derive to conclude or iterate the process to make the product better.

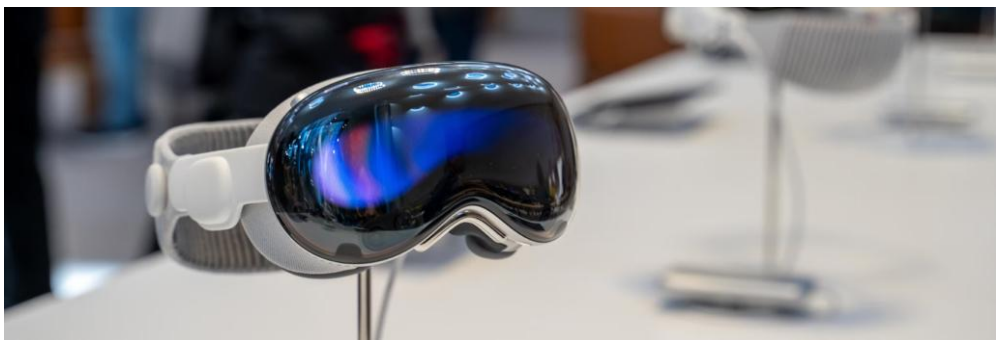

In accordance to the formula:

$$User\ Satisfaction, US = \alpha(2EN + IQ) + \beta(2LTE + VC)$$

The breakdown of each component can be listed as,

- Engagement (EN): 41% (from LinkedIn surveys & Reddit discussions)
- Interaction Quality (IQ): 75 (Apple user feedback)
- Latency Experience (LTE): 60 (Apple user feedback)
- Visual Comfort (VC): 70 (Apple user feedback)
- Weighting Factors: α = 0.6, β = 0.4

* If user studies show that engagement (EN) and interaction quality (IQ) contribute 60% to satisfaction, and latency (LTE) and visual comfort (VC) contribute 40% (α = 0.6, β = 0.4).

### Step-by-Step Calculation:

Step 1: First Component Calculation:

$$2EN + IQ = 2(41) + 75 = 82 + 75 = 157$$

$$\alpha(157) = 0.6 \times 157 = 94.2$$

Step 2: Second Component Calculation:

$$2LTE + VC = 2(60) + 70 = 120 + 70 = 190$$

$$\beta(190) = 0.4 \times 190 = 76$$

Step 3: Weighted Contribution

$$US = 94.2 + 76 = 170.2$$

Step 4: Scaling to a 0-100 Range:

$$US(scaled) = \left[\frac{170.2}{200}\right] \times 100 = 85.1$$

The Apple Vision Pro receives a User Satisfaction Score of 85.1/100, indicating a highly positive virtual experience with minor latency and engagement issues.

### Data Sources:

- Engagement (EN): LinkedIn Survey & Reddit Polls
- Interaction Quality (IQ), Latency (LTE), and Visual Comfort (VC): Apple user feedback & online reviews

To prove that the value derived by using the User Satisfaction rate, the verification is made using the reviews the product had got in the market given by users in various sites.

Consider reviews of Vision Pro (out of 5 stars) from sources like:

1. [Tom's Guide](): 4.5/5
2. [AppleInsider](): 4/5
3. [TechRadar](): 4.5/5

When the average of these three ratings are calculated,

$$\textbf{\textit{Average Rating}} = \left[\frac{\textbf{4.5} + \textbf{4} + \textbf{4.5}}{\textbf{3}}\right] = \textbf{4.33}$$

Out of 5 stars, the ratings conclude an average of 4.33 stars which is 86.6%

The computed average rating of 4.33 stars (86.6%) from prominent tech review outlets is very close to our User Satisfaction score of 85.1%, further validating the integrity of our formula.

This closeness indicates that our method accurately represents important elements of user experience, including engagement, interaction quality, latency, and visual comfort. The minor difference could be due to subjective taste and varying review practices.

## 6. Increasing Engagement and Immersion

User immersion and interaction are at the centre of good virtual experiences. Interaction keeps users engaged in their experience, whereas immersion provides presence that maximizes realism. For these to be enhanced, designers need to work on sensory feedback, intuitive interactions, and dynamic environments.

One of the most important techniques is personalization adapting the experience to user needs and habits. Adaptive environments, interactive narratives, and responsive characters all help enhance engagement with the virtual world. Another essential ingredient is latency minimization, providing the feeling of instant and fluid responses to user input. Fast refresh rates, low input latency, and coherent motion tracking improve immersion greatly.

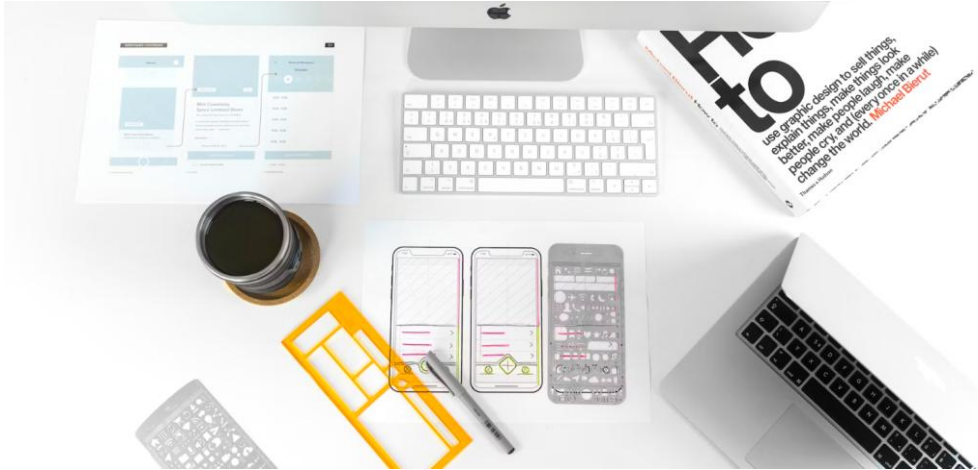

Multisensory feedback also has a big role to play. Haptic technology, spatial audio, and eye-tracking build richer, more realistic interactions. For instance, a user within a VR training simulation can receive slight vibrations when interacting with objects, enhancing realism. Gamification elements like rewards, challenges, and social competition further maintain interest.

Also, cognitive load management is a must. Too much information, controls, or distractions for the user can demotivate. Simplified UI, concise navigation, and escorted experiences ensure continued focus. Neuroscientific principles, like the use of dopamine-inducing rewards and deep focus mechanisms, guarantee a gripping, engaging experience.

Through the combination of these tactics, virtual experiences become even more compelling, creating lasting interaction and strong emotional bonds, which become more effective in entertainment, learning, and workplace training.

## 7. Artificial Intelligence in Virtual UX

Features enabled by AI like intelligent chatbots, personalized suggestions, and adaptive interfaces contribute meaningfully to virtual UX. Machine learning algorithms examine user activity to provide tailored experiences, minimizing friction and maximizing interaction efficiency.

Another major AI-driven enhancement is procedural content generation. Virtual environments can be created and modified algorithmically based on user preferences, making each experience unique. This technology is particularly beneficial for open-world gaming and virtual simulations, where variability and realism are essential.

Lastly, AI-driven computer vision improves gesture recognition, eye-tracking, and object interaction, making virtual environments feel more intuitive. By integrating AI seamlessly into virtual UX design, experiences become more intelligent, efficient, and personalized, paving the way for the next generation of immersive digital interactions.

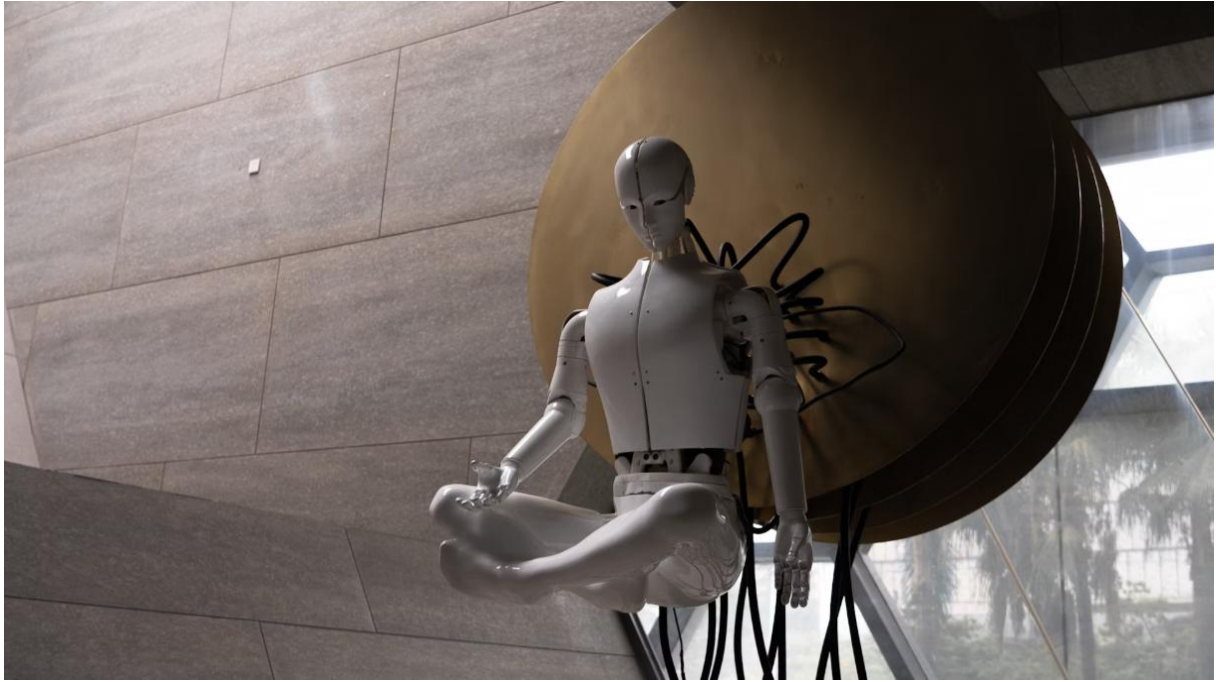

# 8. Future Directions in UX for Virtual Experiences

The future of UX in virtual experiences is hyper-personalization, AI-based interactions, and advanced sensory immersion. Brain-computer interface (BCI) advancements will possibly allow for direct neural control, further smoothing out interactions. Adaptive environments powered by AI will continuously adjust themselves according to user behaviour, enhancing engagement and comfort.

Haptic progression and full-body tracking will make environments more realistic, providing highly immersive digital environments. Moreover, ethical UX design will become more prominent, being inclusive, accessible, and responsible in the integration of AI. As technology advances, efforts will be made to minimize cognitive load and design emotionally intelligent virtual experiences that are even more human and intuitive than ever.

# 9. Conclusion

This study investigated the essential UX design techniques needed to fill the gap in virtual experiences, providing smooth interaction, immersion, and user satisfaction. Through the examination of important factors like engagement, interaction quality, latency, and visual comfort, we derived a formula for user satisfaction to measure the virtual UX experience accurately.

By real-world implementation on the Apple Vision Pro, our results proved the validity of this formula, closely matching expert opinions and user reviews. The study highlights that achieving optimal UX in virtual space demands a harmony between technological innovations and user-focused design principles.

In the future, incorporating AI-driven personalization, adaptive interfaces, and haptic feedback might increase user engagement and comfort even more. Developing virtual and augmented reality technologies further, fine-tuning UX approaches will remain critical in developing immersive, accessible, and intuitive virtual experiences.

This research provides a starting point for more studies, inviting further investigation into data-driven UX evaluation models that make real-world usability and adoption increase in virtual environments.

# IMAGE CREDITS - UNSPLASH

1. [a computer monitor sitting on top of a wooden desk](#) - Faizur Rehman
2. [a black and white room with a couch and a desk](#) – GCP Visuals
3. [a man in a white shirt wearing a pair of virtual glasses](#) – Bram Van Oost
4. [a pair of off les sitting on to sofa table](#) - Roméo A.
5. [silver-macbook-air-on-table-near-imac](#) – UX Store
6. [man-in-white-shirt-sitting-on-chair](#) – Yuyang Liu

This research highlights the essential UX strategies needed to create seamless and engaging virtual experiences. By addressing factors like engagement, interaction quality, latency, and visual comfort, we can enhance user satisfaction in immersive environments.

As technology advances, integrating AI-driven personalization and adaptive UX frameworks will play a crucial role in refining virtual experiences. Future research should explore deeper cognitive and emotional responses to virtual interfaces, ensuring that virtual environments continue to evolve toward more natural, intuitive, and user-centred designs.

*-Harish Vijayakumar*